\documentclass[aip,jcp,reprint,notitlepage,amsmath,amsfonts,amssymb,floatfix]{revtex4-1}
\usepackage{graphicx}
\usepackage{bm}
\usepackage{color}

\begin{document}
\title{Geminal-Based Configuration Interaction}
\author{Thomas M. Henderson}
\affiliation{Department of Chemistry, Rice University, Houston, TX 77005-1892}
\affiliation{Department of Physics and Astronomy, Rice University, Houston, TX 77005-1892}

\author{Gustavo E. Scuseria}
\affiliation{Department of Chemistry, Rice University, Houston, TX 77005-1892}
\affiliation{Department of Physics and Astronomy, Rice University, Houston, TX 77005-1892}
\date{\today}

\begin{abstract}
The antisymmetrized geminal power (AGP) wave function has a long history and considerable conceptual appeal, but in many situations its accuracy is wanting.
Here, we consider a form of configuration interaction (CI) based upon the AGP wave function and taking advantage of its killing operators to construct an excitation manifold.
Our geminal CI reduces to standard single-determinant--based CI in the limit in which AGP reduces to a single determinant.
It substantially improves upon AGP in the reduced BCS Hamiltonian, which serves as a prototype for the kinds of strong pairing correlations relevant in Bardeen-Cooper-Schrieffer--style superconductivity.
Moreover, our geminal CI naturally generalizes to add correlation to more general geminal-based wave functions than AGP.
\end{abstract}

\maketitle

\section{Introduction}
Suppose one has an approximate wave function $|\Phi\rangle$ which one wishes to improve.  A simple route toward doing so is to write a new wave function
\begin{equation}
|\chi\rangle = |\Phi\rangle + \sum c_\mu |\Phi_\mu\rangle
\end{equation}
where the states $|\Phi_\mu\rangle$ are linearly-independent of $|\Phi\rangle$ and the coefficients $c_\mu$ are parameters to be optimized.  By the variational principle, $|\chi\rangle$ almost certainly has lower energy than does $|\Phi\rangle$; at the very least, it cannot be higher.

The trick, of course, is to make a clever choice of states $|\Phi_\mu\rangle$.  We would like $|\Phi_\mu\rangle$ to be linearly independent of $|\Phi\rangle$ and we would like Hamiltonian and matrix elements between $|\Phi\rangle$ and $|\Phi_\mu\rangle$ and between states $|\Phi_\mu\rangle$ and $|\Phi_{\mu^\prime}\rangle$ to be readily evaluated.  If $|\Phi\rangle$ has a symmetry of the exact wave function, we would like $|\Phi_\mu\rangle$ to do likewise.  And of course we would like to choose states $|\Phi_\mu\rangle$ which provide the most improvement to $|\Phi\rangle$ for the least amount of work.

One convenient way to choose the states $|\Phi_\mu\rangle$ is to look for operators $K_\mu(\Phi)$ whose action on $|\Phi\rangle$ is to annihilate it:
\begin{equation}
K_\mu(\Phi) |\Phi\rangle = 0,
\end{equation}
which we shall refer to as killing operators of $|\Phi\rangle$.  Then the subset of all operators $K_\mu^\dagger(\Phi)$ which are not themselves killing operators provides a convenient way to construct the states $|\Phi_\mu\rangle$, which we can create as simply
\begin{equation}
|\Phi_\mu\rangle = K_\mu^\dagger(\Phi) \, |\Phi\rangle.
\end{equation}
These states are automatically orthogonal to $|\Phi\rangle$ (and therefore linearly independent of it).  Further, matrix elements can be readily evaluated because we can adopt a normal order in which all operators $K_\mu^\dagger$ appear adjacent to the bra and all operators $K_\mu$ appear adjacent to the ket; by reordering the various operators in a given matrix element with the aid of commutation or anticommutation rules, we can extract the non-zero portion of the matrix element as that piece which contains neither $K$ nor $K^\dagger$, and we are left with expectation values only with respect to $|\Phi\rangle$ which we presume can be easily computed.  Additionally, if we wish to enforce that the state $|\Phi_\mu\rangle$, like the state $|\Phi\rangle$, is an eigenstate of some symmetry operator $S$ with eigenvalue $s$, then we merely need to further limit our choice of killing operators to those which commute with $S$.  The (hermitian adjoints of) the killing operators thus have many desirable properties which make them well-suited to add corrections to $|\Phi\rangle$.

This observation is by no means new.  Indeed, the standard particle-hole construction in configuration interaction (CI) or coupled cluster theory proceeds entirely along these lines, and similar concerns motivate development in multi-reference theories.  In this manuscript we seek to follow the same path, using the antisymmetrized geminal power (AGP) wave function\cite{Coleman1965} as the reference $|\Phi\rangle$.  To do so, we first need to construct killing operators of the AGP wave function; this, in turn, requires us to say a few words about AGP.

\section{AGP And Its Killing Operators}
The AGP wave function can be written as
\begin{equation}
|\mathrm{AGP}\rangle = \left(\Gamma^\dagger\right)^N |-\rangle
\end{equation}
where $|-\rangle$ is the physical vacuum, $N$ is the number of electron pairs in the system, and $\Gamma^\dagger$ creates a geminal:
\begin{equation}
\Gamma^\dagger = \sum \eta_p \, P_p^\dagger.
\end{equation}
Here, $P_p^\dagger$ creates an electron in level $p$ and another in a level $\bar{p}$ paired with $p$:
\begin{equation}
P_p^\dagger = c^\dagger_p \, c^\dagger_{\bar{p}}.
\end{equation}

The AGP wave function has a symmetry known as seniority in which levels $p$ and $\bar{p}$ are either both occupied or both empty in every determinant in the expansion of the AGP wave function.  General Hamiltonians, of course, do not have seniority as a symmetry, so typically we will need killing operators which do not necessarily conserve seniority.  We shall have more to say on this point later (and indeed, such operators for AGP appeared in the literature many years ago\cite{Weiner1983}), but for now we will limit our attention to seniority-conserving killing operators, which we may construct by making use of the operators $P_p^\dagger$ as well as 
\begin{subequations}
\begin{align}
P_p &= c_{\bar{p}} \, c_p,
\\
N_p &= c^\dagger_p \, c_p + c^\dagger_{\bar{p}} \, c_{\bar{p}}
\end{align}
\end{subequations}
which also preserve seniority.

The only one-body operators we have at hand which preserve both seniority and particle number are the number operators $N_p$; as they are Hermitian, if a linear combination of the $N_p$'s is a killer, so is its adjoint (we assume that the parameters $\eta_p$ and the coefficients in any linear combination of operators intended for use as a killing operator are all real, which guarantees that the wave function also respects complex conjugation symmetry).  Thus, the relevant seniority- and number-conserving killing operators of the AGP wave function are two-body or higher.  We can write these killing operators as
\begin{equation}
K_{pq} = a \, P_p^\dagger \, P_q + b \, P_q^\dagger \, P_p + c \, N_p + d \, N_q + e \, N_p \, N_q
\end{equation}
where we assume $p \ne q$ (for $p = q$, the operator is Hermitian and thus of no use to us).

A simple way to guarantee that $K_{pq}$ is a killing operator is to adjust the coefficients $a$, $b$, $c$, $d$, and $e$ such that $K_{pq}$ commutes with $\Gamma^\dagger$; if this is the case, then
\begin{equation}
K_{pq} |\mathrm{AGP}\rangle = \left(\Gamma^\dagger\right)^N \, K_{pq} |-\rangle
\end{equation}
and as $K_{pq}$ annihilates the physical vacuum, it would then also annihilate the AGP state.

Using the commutation relations
\begin{subequations}
\begin{align}
[P_p,P_q^\dagger] &= \delta_{pq} \, \left(1 - N_p\right),
\\
[N_p,P_q^\dagger] &= 2 \, \delta_{pq} \, P_q^\dagger,
\end{align}
\end{subequations}
one can verify that up to an overall multiplicative factor, the unique operator $K_{pq}$ which has these properties is
\begin{align}
K_{pq} &= \eta_p^2 \, P_p^\dagger \, P_q + \eta_q^2 \, P_q^\dagger \, P_p
\\
 &+ \frac{1}{2} \, \eta_p \, \eta_q \, \left(N_p \, N_q - N_p - N_q\right).
\nonumber
\end{align}
Note that $K_{pq} = K_{qp}$.  For $\eta_p \neq \eta_q$, we have $K_{pq}^\dagger \neq K_{pq}$, so we can hope to write a sort of geminal configuration interaction (GCI) state as
\begin{equation}
|\mathrm{GCI}\rangle = |\mathrm{AGP}\rangle + \sum_{\substack{p > q\\ \eta_p \neq \eta_q}} c_{pq} \, K_{pq}^\dagger \, |\mathrm{AGP}\rangle.
\end{equation}

\begin{figure}[t]
\includegraphics[width=\columnwidth]{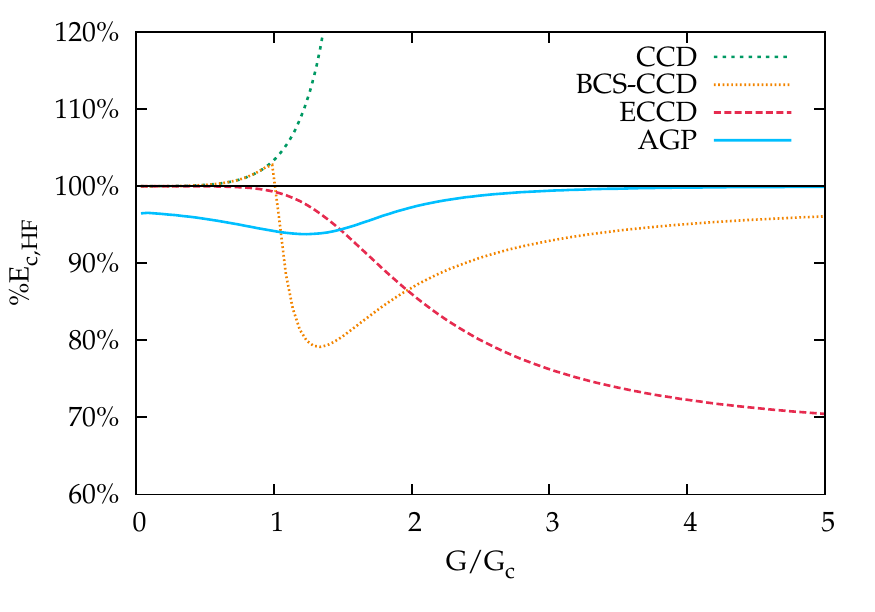}
\caption{
Percentage of correlation energy with respect to Hartree-Fock recovered by coupled cluster doubles (CCD), broken-symmetry CCD (BCS-CCD), extended CCD (ECCD), and AGP in the half-filled, 16-site pairing Hamiltonian as a function of $G/G_c$, where $G_c$ is the value of $G$ at which the mean-field spontaneously breaks number symmetry.
\label{Fig:ConventionalMethods}}
\end{figure}

This is the wave function we intend to investigate in this work.  We will limit our attention to AGP wave functions which respect spin symmetry, so that the orbitals $p$ and $\bar{p}$ are the $\uparrow$-spin and $\downarrow$-spin spinorbitals for the same spatial orbital.  Further, because we have chosen to consider only the seniority-conserving kiling operators, we should limit ourselves to Hamiltonians for which seniority is a symmetry; accordingly, we will specialize to the pairing or reduced BCS Hamiltonian, which can be written entirely in terms of the seniority zero generators $P$, $P^\dagger$, and $N$ and for which the exact energy can be obtained via a form of Bethe ansatz which simply requires solving the nonlinear Richardson equations.\cite{Richardson1963,Dukelsky2004}  Note that even for general Hamiltonians, only the portion of the Hamiltonian which itself conserves seniority contributes to AGP expectation values involving $H$, $K$, and $K^\dagger$.

\section{The Pairing Hamiltonian}
The pairing Hamiltonian can be written as
\begin{equation}
H = \sum_p \epsilon_p \, N_p - G \, \sum_{pq} P_p^\dagger \, P_q.
\end{equation}
While other choices are possible, we take $\epsilon_p = p$ and $G>0$.  Note that for positive $G$ the Hamiltonian has an attractive two-body interacton rather than the more familiar repulsive two-body interaction.  The mean-field symmetry which spontaneously breaks is, accordingly, number symmetry rather than spin symmetry, giving rise to the well-known Bardeen-Cooper-Schrieffer (BCS) wave function.\cite{Bardeen1957}  Note also that the interaction is infinite in range.

This Hamiltonian was introduced as a phenomenological model to describe pairing correlations in which electron pairs interact with the hole pairs they have left behind, but has also been used to model ultrasmall superconducting grains.\cite{Braun1998,Dukelsky1999,Sierra2000}

\begin{figure*}[t]
\includegraphics[width=\columnwidth]{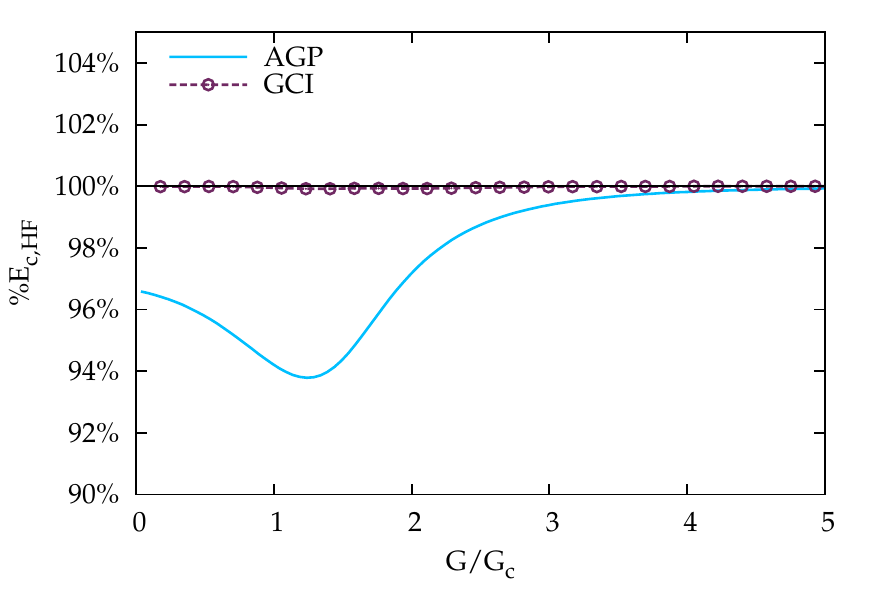}
\hfill
\includegraphics[width=\columnwidth]{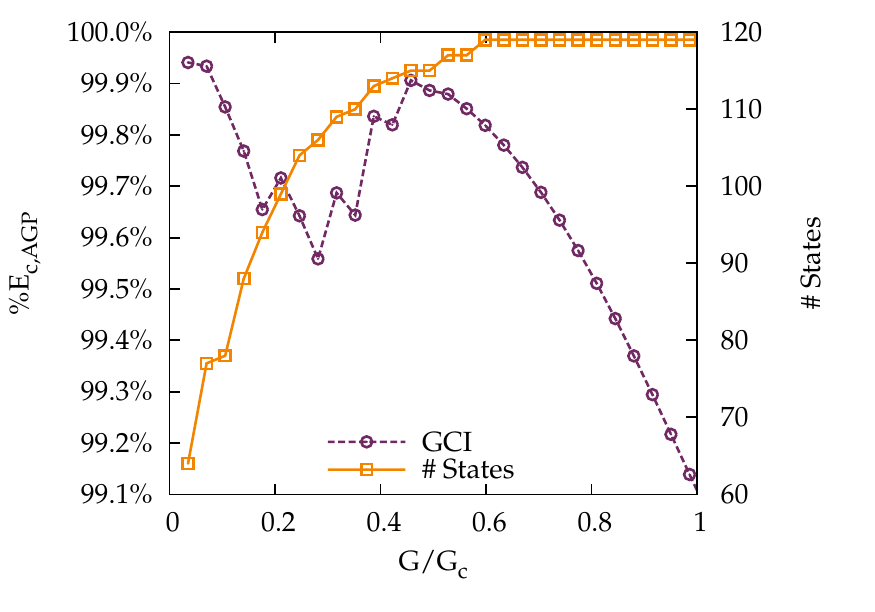}
\caption{
Left panel:  Percentage of correlation energy with respect to Hartree-Fock recovered in the half-filled, 16-site pairing Hamiltonian as a function of $G/G_c$.
Right panel: Percentage of correlation energy with respect to AGP recovered, and the number of additional states created by the killing operators.
\label{Fig:16Sites8Pairs}}
\end{figure*}

Remarkably, the Hamiltonian can be solved exactly.  Its eigenstates can be written as
\begin{subequations}
\begin{align}
|\Psi\rangle &= \prod \Gamma_\mu^\dagger |-\rangle,
\\
\Gamma_\mu^\dagger &= \sum_p \frac{1}{2 \, \epsilon_p - R_\mu} \, P_p^\dagger,
\end{align}
\end{subequations}
where the energy is given by the sum of the rapidities $R_\mu$, which satisfy a set of coupled nonlinear equations.  Note that the exact wave function is simply an antisymmetric product of non-orthogonal geminals.

Although the exact solution is readily available, this Hamiltonian is very difficult for conventional electronic structure methods such as single-reference coupled cluster theory\cite{Dukelsky2003,Henderson2014} and even extended coupled cluster.\cite{Henderson2015}  This can readily be seen by Fig. \ref{Fig:ConventionalMethods}, which shows the fraction
\begin{equation}
\%E_{c,\mathrm{HF}} = \frac{E - E_\mathrm{HF}}{E_\mathrm{exact} - E_\mathrm{HF}}
\end{equation}
of the exact correlation energy with respect to Hartree-Fock recovered as a function of $G/G_c$. where $G_c$ refers to the value of $G$ at which the mean-field undergoes spontaneous symmetry breaking toward a number-broken mean-field solution.  But while sophisticated coupled cluster approaches fail for this problem, AGP provides good albeit imperfect results for almost all $G$ and is an eigenstate in the $G \to \infty$ limit.\footnote{The manner of plotting in Fig. \ref{Fig:ConventionalMethods} suggests errors in AGP for small $G$, but in fact AGP is exact for $G \to 0$; Fig. \ref{Fig:ConventionalMethods} merely reflects the fact that AGP delivers the wrong $\mathcal{O}(G^2)$ contribution to the total energy}  Note that in the repulsive case the situation is very different, and traditional coupled cluster performs exceptionally well.  The attractive pairing Hamiltonian thus forms an interesting test case for our geminal CI.  Our hope is that if we can correctly describe this exactly solvable model, we can also correctly describe related but more physical problems for which no exact solution exists but which pose similar computational challenges.

\section{Results}
To test the accuracy of our geminal CI ansatz, we return to the half-filled 16-site pairing Hamiltonian, which is shown in Fig. \ref{Fig:16Sites8Pairs}, the left panel of which can be directly compared to Fig. \ref{Fig:ConventionalMethods}.  Although the geminal CI is not exact, it has successfully remedied almost all of the deficiencies of AGP.  This is made much clearer by examination of the right panel of Fig. \ref{Fig:16Sites8Pairs}, which shows the fraction $\%E_{c,\mathrm{AGP}}$ of correlation with respect to AGP recovered by the geminal CI wave function and shows that GCI removes, in this case, more than 99\% of the energy error made by AGP.

The right panel also reveals a curious (and unpleasant) feature of the geminal CI.  While for small $G$ the AGP reference becomes more like a single determinant (see below) for which we expect the number of double excitations to be $N \, (M-N)$ where $N$ is the number of pairs and $M$ is the number of levels, the number of killing operators we construct is $M \, (M-1)/2$.  Several must be linearly-dependent for small $G$, and these linearly-dependent states must be removed from the diagonalization.  This, we expected.  What was unexpected is that we always encounter one linearly-dependent mode even for larger $G$, and this appears to be true independent of the number of pairs $N$ and number of levels $M$.  Put differently, there always appears to be at least one linear combination of operators $K_{pq}^\dagger$ which is a killing operator of AGP and which must therefore be excluded.  We do not at present have a better prescription for finding this mode than by diagonalizing the metric $\langle K_{pq} \, K_{rs}^\dagger \rangle$ and eliminating the mode with zero eigenvalue.  Note that this zero eigenvalue is not simply a small eigenvalue, but rather one which vanishes to the double precision in which we have implemented our geminal CI.

\begin{figure*}[t]
\includegraphics[width=\columnwidth]{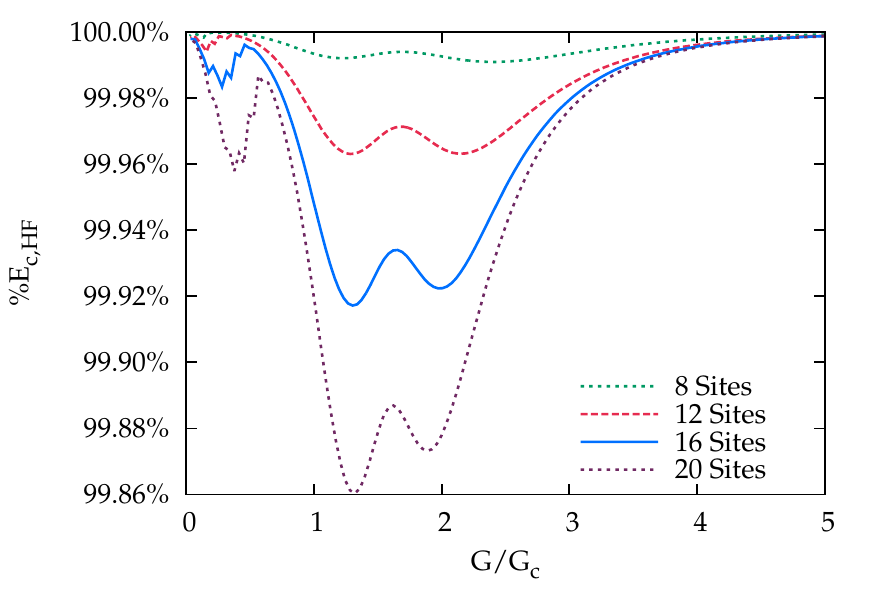}
\hfill
\includegraphics[width=\columnwidth]{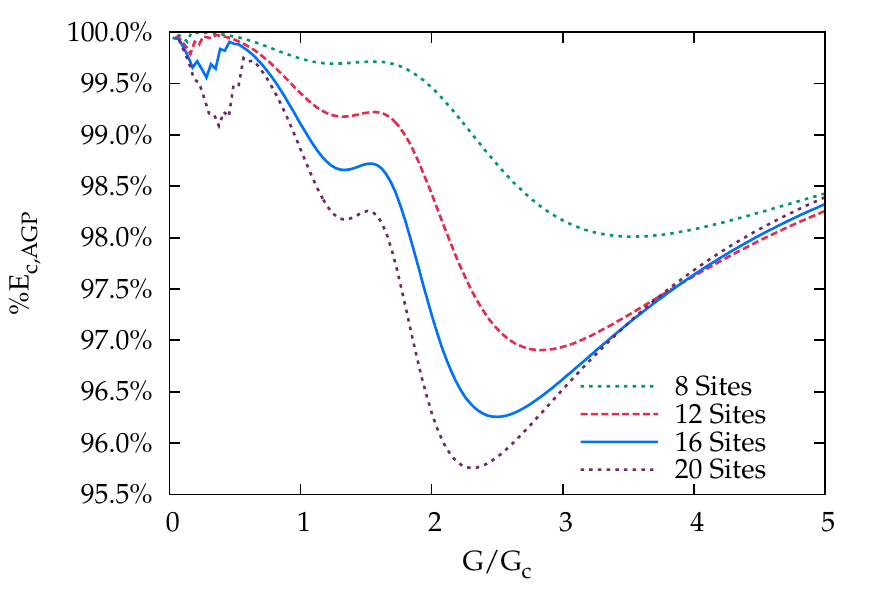}
\caption{
Percentage of correlation energy recovered in the half-filled pairing Hamiltonian for several different system sizes.
Left panel: correlation energy with respect to Hartree-Fock.
Right panel: correlation energy recovered with respect to AGP.
\label{Fig:SystemSize}}
\end{figure*}

To get a sense of how our method performs as we adjust system size, Fig. \ref{Fig:SystemSize} shows the fraction of correlation energy recovered in several different pairing Hamiltonians.  Clearly our results deteriorate somewhat as system size increases, but the accuracy remains uniformly excellent.  Results away from half-filling (not shown) are similar.  Neither AGP nor truncated CI are fully extensive methods, and we ought not expect geminal CI to be extensive either.  The pairing Hamiltonian, unfortunately, is not a good place to test the error in extensivity, because the exact energy is not linear in system size in the first place since the interaction has infinite range.  Nonetheless, as the right panel of Fig. \ref{Fig:SystemSize} makes clear, a large portion of the size-dependence in the geminal CI is due to the underlying AGP rather than to the geminal CI \textit{per se}.

\begin{figure}[b]
\includegraphics[width=\columnwidth]{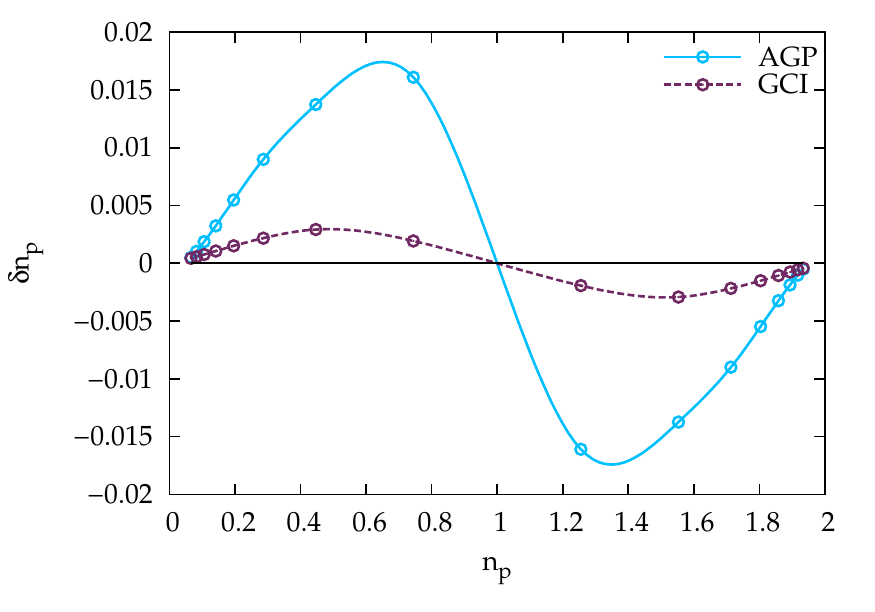}
\caption{Errors $\delta n_p$ in the occupation numbers $n_p$ for the 16-site, 1/2-filled pairing model at $G \sim 2 \, G_c$.
\label{Fig:OccNums}}
\end{figure}

Although obtaining accurate energies is perhaps the chief priority, we would also like similar accuracy for the calculation of properties (or, in other words, we would like to get both the energy and the wave function right).  To assess this, Fig. \ref{Fig:OccNums} shows the errors in the occupation numbers $n_p = \langle N_p \rangle$ for the 16-site, half-filled pairing Hamiltonian at $G \sim 2 \, G_c$.  This value is large enough that we expect to see occupations differ appreciably from 0 and 2 (the mean-field result) without being too large.  Clearly the geminal CI significantly reduces the error in the occupation numbers, and thus in typical single-particle properties.  We should note that the plot has a symmetry around $n_p = 1$ which is just a consequence of a symmetry of the half-filled pairing Hamiltonian.\cite{Hirsch2002}

\section{Discussion}
We have seen that for the pairing Hamiltonian, our geminal CI is remarkably accurate.  Here, we wish to discuss the extension to more realistic Hamiltonians which do not have seniority as a symmetry.

For such Hamiltonians, the relevant killing operators are one-body in nature, and can be written as
\begin{equation}
D_{pq} = \eta_p \, E^p_q - \eta_q \, E^q_p
\end{equation}
for $p \neq q$ in terms of operators
\begin{equation}
E^p_q = c_p^\dagger \, c_q + c_{\bar{p}}^\dagger \, c_{\bar{q}}
\end{equation}
which become the standard spin-integrated unitary group generators when levels $p$ and $\bar{p}$ are the $\uparrow$- and $\downarrow$-spinorbitals corresponding to the same spatial orbital.  Killers of AGP were first discussed in quantum chemistry,\cite{Weiner1983} and recently applied by Dukelsky and collaborators to superfluid nuclei,\cite{Dukelsky2019} albeit not in the same $D_{pq}$ form as above.  We may think of the operators $D$, $D^\dagger$, and $N$ as a basis of one-body operators and so in principle can use them to write a normal order of the form $D^\dagger \, N \, D$.

Our killing operators $K_{pq}$ are essentially just $1/2 \, D_{pq}^2$; while not strictly identical, $K_{pq}$ and $1/2 \, D_{pq}^2$ act identically on seniority-zero states.  For problems such as the molecular Hamiltonian, we can generalize our geminal CI to the form
\begin{align}
|\mathrm{GCI}\rangle = \big(1 &+ \sum c_{pq} \, D_{pq}^\dagger
\\
 &+ \frac{1}{2} \, \sum c_{pqrs} \, \{D_{pq}^\dagger,D_{rs}^\dagger \} + \ldots \big) |\mathrm{AGP}\rangle.
\nonumber
\end{align}
The expectation value of the Hamiltonian with respect to this wave function can readily be taken and minimized with respect to the variational parameters, and we shall report on its applications in due time.  The anticommutator in the quadratic term is present to ensure that it corresponds to a genuinely two-body operator (because the commutator of $D_{pq}^\dagger$ with $D^\dagger_{rs}$ is a non-zero one-body operator).  

Let us say a few words about the single-determinant limit.  If we have only $N$ non-zero parameters $\eta_p$, then the AGP wave function reduces to a single determinant.  Using the conventional notation in which $a$ and $b$ index unoccupied orbitals with $\eta = 0$ and $i$ and $j$ index occupied orbitals with $\eta \neq 0$, then
\begin{subequations}
\begin{align}
D_{ab} &= 0,
\\
D_{ia} &= \eta_i \, E^i_a,
\\
D_{ij} &= \eta_i \, E^i_j - \eta_j \, E^j_i.
\end{align}
\end{subequations}
The adjoint operators $D_{ij}^\dagger$ annihilate the reference determinant because $i \neq j$, while the adjoint operators $D_{ia}^\dagger$ are, up to an overall factor, the usual particle-hole excitation operators.  Thus, in the single-determinant limit of AGP, our geminal CI reduces to standard single-reference configuration interaction when using the $D_{pq}$ killers (and to the CI analog of pair coupled cluster doubles when using the $K_{pq}$ killers).

\begin{figure}[t]
\includegraphics[width=\columnwidth]{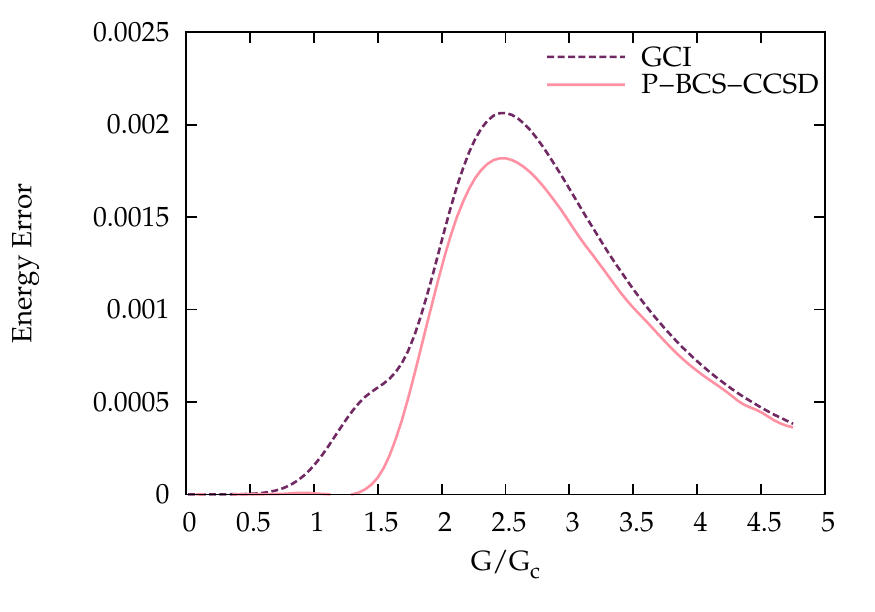}
\caption{Total energy errors in the half-filled, 12-site pairing Hamiltonian with geminal CI and with symmetry-projected coupled cluster theory.
\label{Fig:CompareEthan}}
\end{figure}

Our geminal CI bears close resemblance to some methods which merge correlation and symmetry projection, essentially because AGP is a symmetry-projected mean-field method (AGP is equivalent to number-projected BCS).  Although the two methods are distinct, geminal CI is particularly closely related to the projected configuration interaction of Ripoche \textit{et al.}\cite{Ripoche2017} (and also to a spin-projected analog due to Tshuchimochi and Ten-no\cite{Tsuchimochi2016a,Tsuchimochi2016b}).  A coupled-cluster generalization of our geminal CI would likewise approximate symmetry-projected coupled cluster theory.\cite{Duguet2014,Duguet2017,Tsuchimochi2017,Qiu2017,Qiu2019}  These methods write a correlated wave function which uses the killing operators of BCS and then act a number projector on the correlated state.  In contrast, our method works directly with those operators which annihilate the \textit{projected} mean-field state.  Figure \ref{Fig:CompareEthan} shows that for a small pairing model, geminal CI provides accuracy similar to symmetry-projected coupled cluster with singles and doubles without needing to worry about the highly technical details of how one might carry out the symmetry projection of a correlated state.

Finally, a word on the name ``geminal CI''.  We have chosen this name because while we have considered only the AGP case, the ideas discussed in this manuscript generalize to other geminal wave functions.  Consider, for example, a geminal wave function
\begin{equation}
|\Psi\rangle = \prod \left(\Gamma_\mu^\dagger\right)^{N_\mu} |-\rangle
\end{equation}
where the geminals created by the different $\Gamma_\mu^\dagger$ are strongly orthogonal (i.e. are expanded in mutually orthogonal single-particle orbital spaces):
\begin{equation}
\Gamma_\mu^\dagger = \sum_{p \in \mu} \eta_p \, P_p^\dagger.
\end{equation}
This geminal wave function is a kind of cluster mean-field state.\cite{JimenezHoyos2015}  It includes AGP on the one hand (when there is a single geminal so that $\prod \left(\Gamma_\mu^\dagger\right)^{N_\mu} = \left(\Gamma^\dagger\right)^N$) and the antisymmetrized product of strongly orthogonal geminals\cite{Surjan1999,Rassolov2002,Surjan2012,Limacher2013,Pernal2014} (APSG) on the other (when the various $N_\mu$ are all equal to 1).  Such wave functions are annihilated by our killing operators $K_{pq}$ and $D_{pq}$ and thus form a perfectly valid reference state for our geminal CI.  An APSG-based version of our geminal CI would be related to both K\'allay and Surj\'an's APSG-based CI method\cite{Kallay1999} and Goddard's generalized valence bond CI.\cite{Carter1988,Faglioni1999}  While we have limited our attention in this manuscript to the AGP case, the theory is considerably more general.

\begin{acknowledgments}
This work was supported by the U.S. National Science Foundation under Grant CHE-1762320. G.E.S. is a Welch Foundation Chair (C-0036).  We thank Jorge Dukelsky for providing his AGP code and Armin Khamoshi for providing code to evaluate AGP density matrices.
\end{acknowledgments}

\bibliography{AGPCI}
\end{document}